\documentclass[showpacs,prb, twocolumn]{revtex4-1}
\usepackage{graphicx}
\usepackage{amsmath}

\begin{document}

\title{Defects-driven magnetism in bulk $\alpha$-Li$ _{3}$N }

\author{Saima Kanwal}

\affiliation{Department of Physics, Quaid-i-Azam University, Islamabad 45320, Pakistan}
\author{Gul Rahman}
\email[Corresponding author: ]{gulrahman@qau.edu.pk}

\affiliation{Department of Physics, Quaid-i-Azam University, Islamabad 45320, Pakistan}
\date{\today}

\begin{abstract}
\textit{Ab-initio} calculations based on density functional theory  with local spin density approximation are used to study defects-driven magnetism in bulk $\alpha$-Li$ _{3}$N. 
Our calculations show that bulk Li$ _{3} $N is a non-magnetic semiconductor.
Two types of Li vacancies (Li-I and Li-II) are considered, and   
Li-vacancies (either Li-I or Li-II type) can induce magnetism in Li$ _{3}$N with a total magnetic moment of 1.0 $\mu_{\rm B}$ which arises mainly due to partially occupied N-$p$-orbitals around the Li vacancies.
The defect formation energies dictate that Li-II vacancy, which is in the Li$ _{2}$N plane, is thermodynamically more stable as compared with Li-I vacancy. The electronic structures of Li-vacancies show half-metallic behavior.
On the other hand N-vacancy does not induce magnetism and has a larger formation energy than Li-vacancies. N vacancy derived bands at the Fermi energy are mainly contributed by the Li atoms. Carbon is also doped at Li-I and Li-II sites, and it is expected that doping C at Li-I site is thermodynamically more stable as compared with Li-II site.
Carbon can induce metallicity with zero magnetic moment when doped at Li-I site, whereas magnetism is observed when Li-II site is occupied by the C impurity atom and C-driven magnetism is spread over the N atoms as well. Carbon can also induce half-metallic magnetism when doped at N site in Li$ _{3}$N, and has a smaller defect formation energy as compared with Li-II site doping. The ferromagnetic (FM) and antiferromagnetic (AFM) coupling between the C atoms is also investigated, and we conclude that FM state is more stable than the AFM state.

\end{abstract}


\maketitle

\newpage

\section{Introduction}
Progression in the spintronics paradigm,\cite{spintro2000,spintro1,spintro2} exploiting both the charge and the spin degree of freedom of an electron, has grabbed an extended attention. In this context, many diluted magnetic semiconductors (DMS) have been discovered in the past to exhibit room temperature (RT) ferromagnetism when doped with transition metals (TM).\cite{TM1,TM2,TM3} However, the controversial origin of magnetism in these materials and the observation of magnetic clusters or secondary phases \cite{MCl1,MCl2} limit  their functioning for practical applications. Intrinsic defects, like vacancies, have been significant for magnetism in semiconductors, \cite{g1,g2,CaO} having different origins of ferromagnetism owing to different crystal environment and local symmetry. Doping a nonmagnetic semiconductor with  nonmagnetic impurity atoms, generally the light elements such as C, N, and Li \cite{SK1,LE2,LE4,LE5} has been found as an alternative to TM doped semiconductors. In this quest, several light element-doped oxides, nitrides and sulphides have been reported to display intrinsic ferromagnetism, where the $p$-orbitals of the impurity atoms play a crucial role in deriving magnetism in the host material and can also be expected to form an impurity band in the bandgap including the Fermi energy ($ E_{\rm F} $) of the otherwise nonmagnetic semiconductor matrix.\cite{K.Li, L.Shen, H.Pan,G.R,S.W.Fan} In addition to induce magnetism, light elements have also been observed  to stabilize the intrinsic defects in the host material by lowering their formation energies.\cite{stb1,stb2} Such DMSs have a considerable magnetic moment and their Curie temperature $T_{c}$ is well above the RT, thus demonstrating their viability for spintronics.

Lithium nitride (Li$ _{3} $N) can be another possible candidate for DMSs which exists in three polymorphs, the hexagonal $\alpha$-Li$ _{3} $N which is stable at room temperature and pressure and is synthesized from the elements at elevated temperature and ambient conditions, the hexagonal $\beta$-Li$ _{3} $N which is obtained from the $\alpha$ form and is stable at moderate pressure (4.2 kbar and 300 K) and cubic $\gamma$-Li$ _{3} $N which transforms from the $\beta$ phase and remains stable upto 200 GPa.\cite{A.L,A.L2}  Li$ _{3} $N has an exceptionally high Li ion conductivity due to intrinsic defects which results in Li-ion hoping from one Li site (occupied) to another (unoccupied).\cite{H} It is used in hydrogen storage battery technology due to its high theoretical H$ _{2} $ capacity,\cite{Hcap1,Hcap2} and is also a component used in the synthesis of nanophase GaN.\cite{GaN} In the past Li$ _{3} $N also served as a host for ferromagnetism induced by TM like Mn, Fe, Co and Ni.\cite{LN5,LN6,LN7} Li vacancies in Li$ _{3} $N are also found to be magnetic.\cite{ostlin} Very recently we also observed magnetism in the Li$ _{2} $N monolayer without any crystal defects.\cite{GR2017}

\begin{figure}[t]
\includegraphics[width=0.2\textwidth]{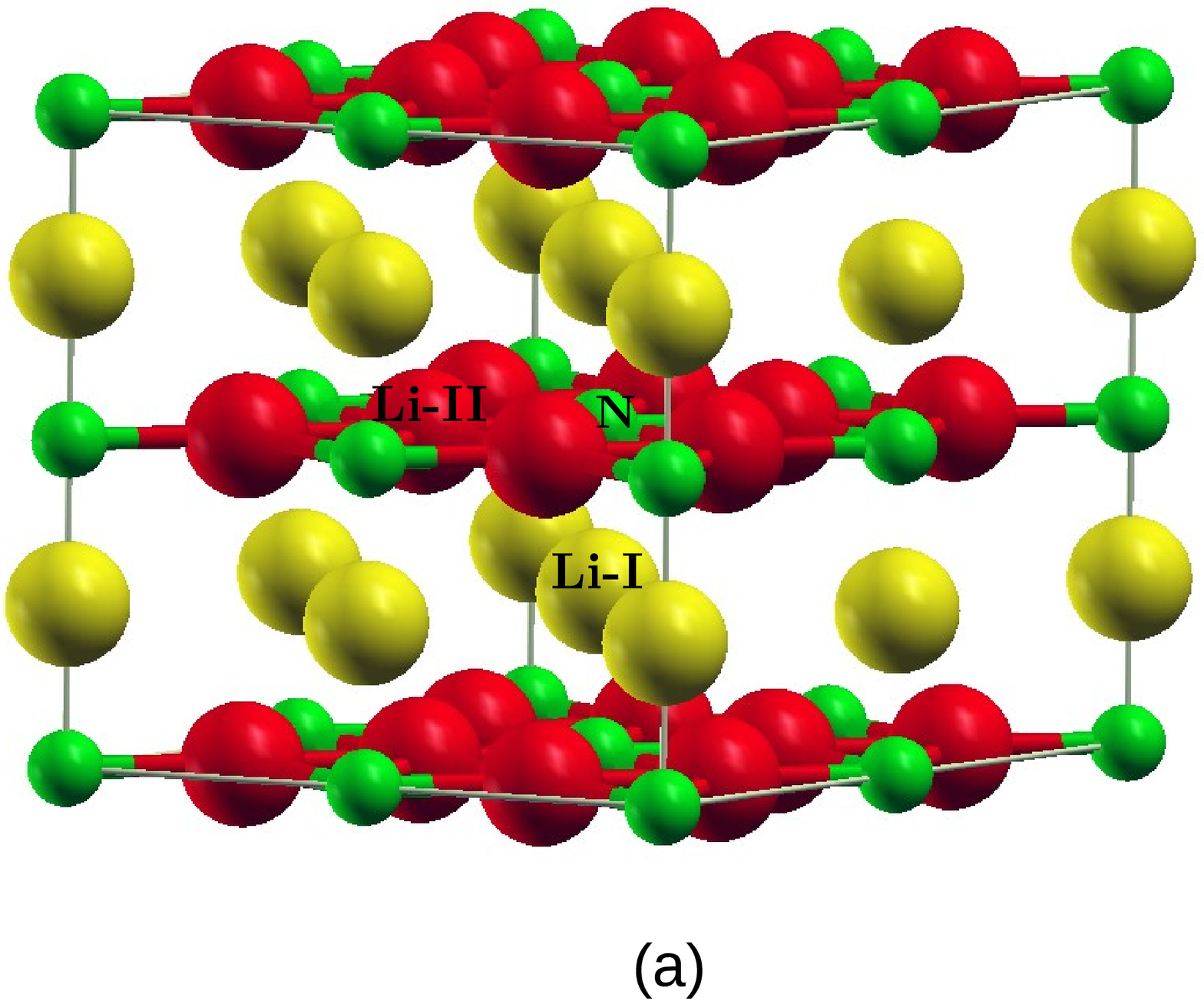}
\includegraphics[width=0.15\textwidth]{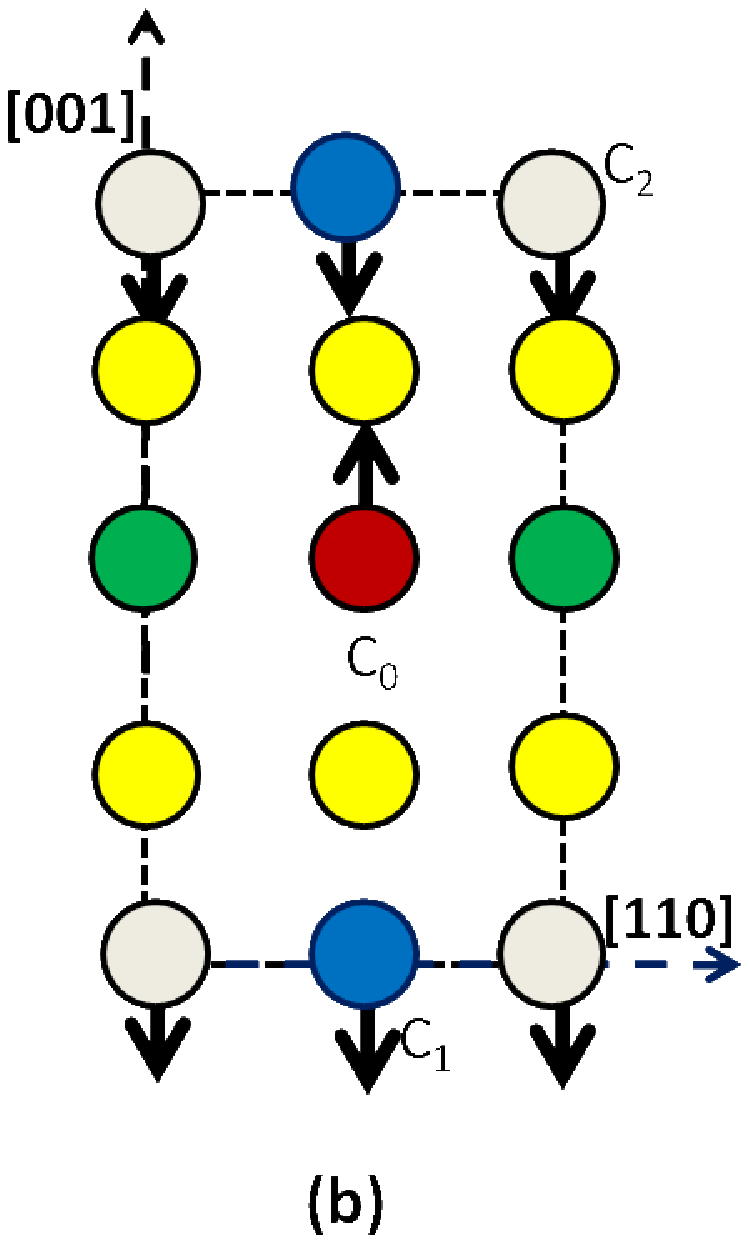}
\caption{ (Color online) (a) Crystal structure of a $2 \times 2 \times 2$ supercell of $ \alpha $-Li$ _{3} $N, consisting of 32 atoms. Yellow and red  balls represent Li-I and Li-II atoms, respectively,  whereas green balls show the N atoms. Labels indicate the atomic sites where  defects are introduced.(b) Magnetic coupling between the C atoms in the (110) plane. C$_{0}$ (brown ball) is the central C atom doped at N site, whereas C$_{1}$ atoms (blue balls) show the case when C$_{0}$--C$_{1}$ is 3.63\,\AA. C$_{2}$ atoms (grey balls) represent the case when C$_{0}$--C$_{2}$ is 4.98\,\AA. Yellow (green) balls represent the Li-I(N) atoms. Arrows indicate the direction of spin-magnetic moment in AFM configuration.  }
\label{str}
\end{figure}

$\alpha$-Li$ _{3}$N has a unique hexagonal crystal structure with four atoms per unit cell, at ambient conditions and equilibrium pressure [Fig.~\ref{str}]. The unit cell parameters are $a$ = 3.648 {\AA} and $c$ = 3.874 {\AA}  with the symmetry point group of $D ^{1} 6h$ (space group $P6/mmm$).\cite{H.J.Beister} $\alpha$-Li$ _{3} $N has a layered structure, consisting of Li$ _{2} $N layers which are widely separated by a pure Li-atoms layers, where Li-atoms occupy a site between the N atoms in the adjacent layers. 
There are two types of Li atoms, denoted as Li-I and Li-II, in Li$ _{3} $N.
Li-I atoms
occupy the $1b$ Wyckoff positions ($x = 0, y = 0, z = 1/2$), whereas
Li-II atoms occupy the $2c$ positions ($x = 1/3, y = 2/3, z = 0$), and the
N atoms are at the $1a$ positions ($x = 0, y = 0, z = 0$). The
Li$ _{2} $N layer is formed by Li-II in the $ab$ plane with edge-shared
N-Li$_{6}$ hexagons and Li-I positions between layers
to form continuous {Li-I-N-Li-I} chains along the $c$ axis.
In Li$ _{3} $N the N atoms have eight Li atoms as
nearest neighbors in which two atoms are at 1.94\AA\, distance along the $c$
axis and six Li atoms at 2.10\AA\, in the $ab$ plane. 
Due to different coordination number and bond lengths in Li$ _{3} $N, it is expected that Li$ _{3} $N will have different properties induced by defects at Li-I and Li-II sites.
X-ray diffraction and powder neutron diffraction studies also reported
1-2\% vacancies in the Li-II position at room temperature,\cite{expRT} and the concentration of vacancy can go up to 4\% at high temperatures. Therefore, it is very important to investigate the electronic structure, thermodynamics, and possible magnetism driven by defects at different lattice sites (Li-I, Li-II, N) in  Li$ _{3} $N and propose a new DMS system based on light elements (Li, C, N) instead of TM.

\section{Computational Method and Models} 

We performed \textit{ab-initio} calculations in the framework of density functional theory (DFT)\cite{DFT, HK} using linear combination of atomic orbitals (LCAO) as implemented in the SIESTA code.\cite{siesta} The local density approximation (LDA)\cite{lda} was adopted to deal with the exchange-correlation functional.  A cutoff energy of 200 Ry for the real-space grid was adopted and for the Brillouin zone sampling a $16\times 16\times 8$ Monkhorst-Pack (MP) $k$-point grid was used for the unit cell and a $6\times 6\times 3$ MP $k$-point grid was used for the $2\times 2\times 2$ supercell in the electronic structure calculations. We employed standard norm-conserving pseudopotentials in their fully nonlocal form. {For the
pseudopotentials, we used a Troullier–Martins \cite{Martin}
form for Li ($2s^{1}2p^{0}3d^{0}$) and N ($2s^{2}2p^{3}3d^{0}$).} A double-$\zeta$ polarized (DZP) basis set for all atoms are employed. The atomic positions were optimized, using the conjugate-gradient algorithm, until the residual Hellmann-Feynman forces were reduced to less than 0.05 eV/{\AA}. Note that convergence with respect to $k$-point sampling and cutoff energy was carefully checked. Test calculations were also carried out using generalized gradient approximation (GGA).\cite{gga}{The LDA/GGA Hamiltonian was solved using the diagonalization method as implemented in the SIESTA code.}

We considered different types of defects including the Li and N vacancies and C doping at various atomic sites to investigate magnetism in the bulk Li$ _{3} $N using a $2\times 2\times 2$ supercell of the primitive cell of Li$ _{3} $N. Since the formation energy (E$ _{\rm f} $) is the cost of inducing defects in a crystal,  and the  concentration of dopant and vacancies in  a crystal depend upon its formation energies, therefore we determined the formation energies for systems with intrinsic defects, i.e., Li or N vacancy as follows,

\begin{equation}
E_{\rm f} = E^{\rm d} - E^{\rm p} +  n \mu_{\rm x}
\label{vac}
\end{equation}

where $\mu_{\rm x}$ is the chemical potential of x (x= Li, N), $n$ represents the number of atoms removed from the system, $ E^{\rm d} $ is the total energy of Li$ _{3} $N with a single Li or N vacancy per supercell. Here, it is important to mention that we created two types of Li vacancies (Li-I and Li-II), therefore $E^{\rm d}$ is different for each case. $ E^{\rm p} $ represents the total energy of pure Li$ _{3} $N  supercell. The chemical potential of Li ($\mu_{\rm Li}$) is calculated as, $\mu_{\rm Li}$ = E (Li), where E (Li) is the total energy per atom of BCC Li and that of N is calculated as $\mu_{\rm N}$ = $\frac{1}{2}$ [E (N$ _{2} $)], where E(N$ _{2} $) is the total energy  of N-molecule.

The formation energy of the C-doped systems is calculated as, 

\begin{equation}
E_{\rm f} = \frac{1}{n}(E^{\rm d} - E^{\rm p} +  n \mu_{\rm x} -  m \mu_{\rm c})
\label{C-doped}
\end{equation}

where $\mu_{\rm x}$  and $\mu_{\rm c}$ are the chemical potentials of  x(x = Li, N) and C respectively,  $n$ shows the number of atoms removed from the system, $m$ is the number of atoms added to the system,  and  $E^{\rm d}$ is the total energy of the corresponding C-doped systems.  The chemical potential of C is calculated as, $\mu_{\rm C}$ = E (C), where E (C) is the total energy per atom of the diamond C.
The formation enthalpy $\Delta H_{f}$ of bulk Li$_{3}$N is calculated as $\Delta H_{f}=\mu_{\rm{Li_{3}N}}-3\mu_{\rm{Li}}-\mu_{\rm{N}}$, where $\mu_{\rm{Li_{3}N}}$ is the total energy of bulk $\alpha$-Li$_{3}$N. The formation energies of all defects were calculated under both Li-rich (N-poor) and N-rich (Li-poor) conditions following our previous approach.\cite{stb1}
Note that in the present manuscript we will only present our LSDA data, and GGA data will be presented for comparison  wherever it is required. {In the above equations, negative formation energy means that the system is thermodynamically more stable as compared with positive formation energy.} 

\section{Results and Discussions}
\begin{figure}[]
\includegraphics[width=0.1\textwidth]{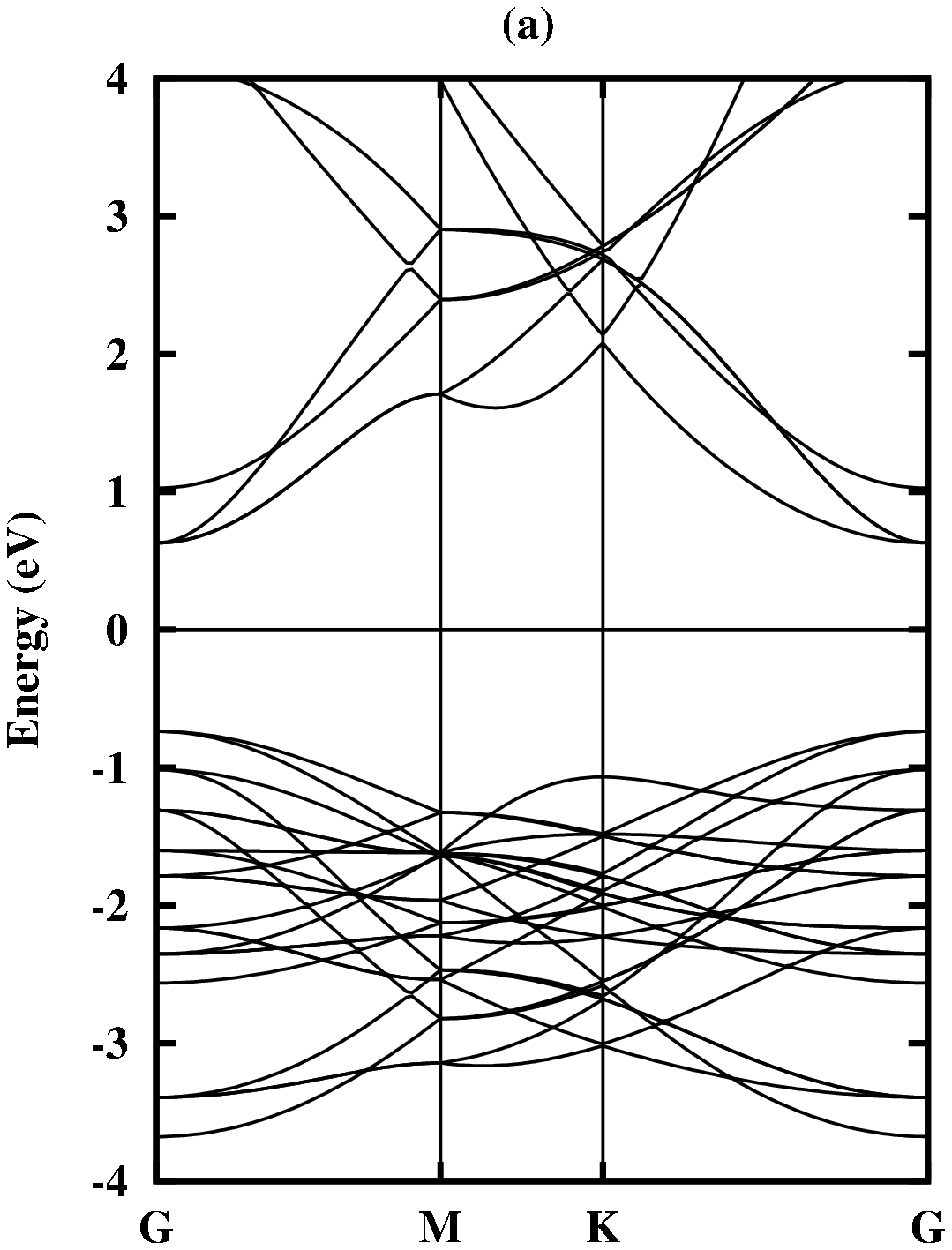}  
\includegraphics[width=0.1\textwidth]{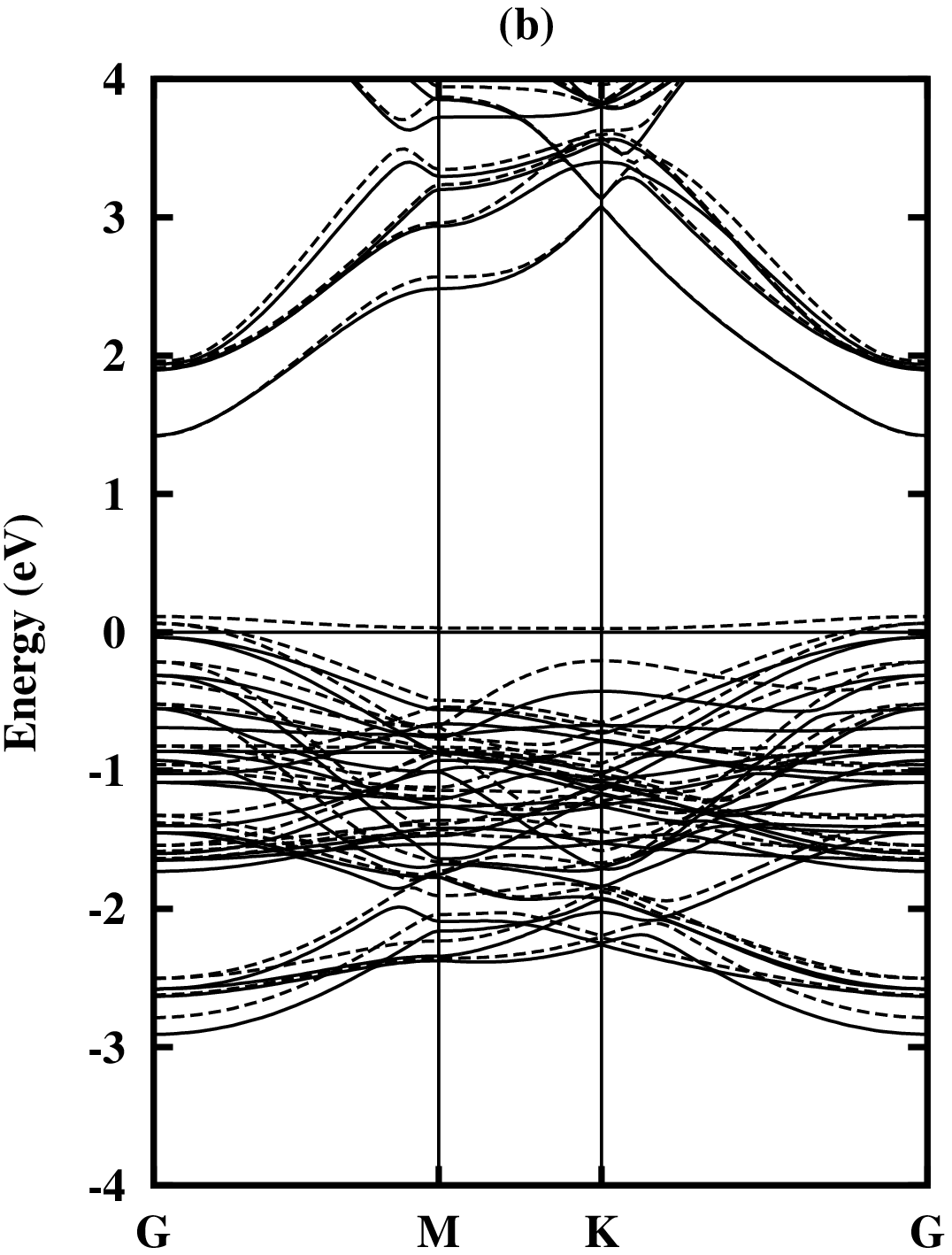}  
\includegraphics[width=0.1\textwidth]{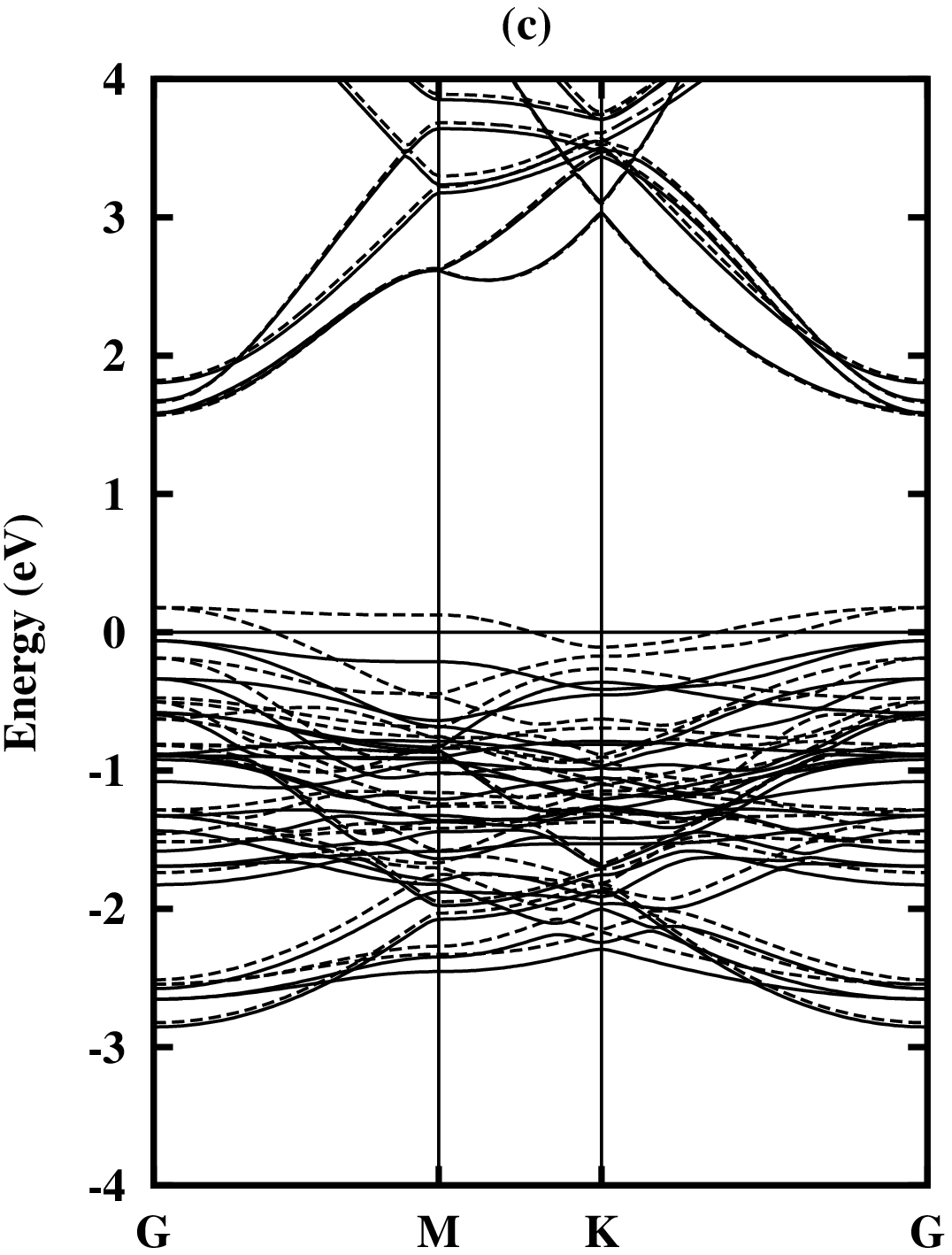}
\includegraphics[width=0.1\textwidth]{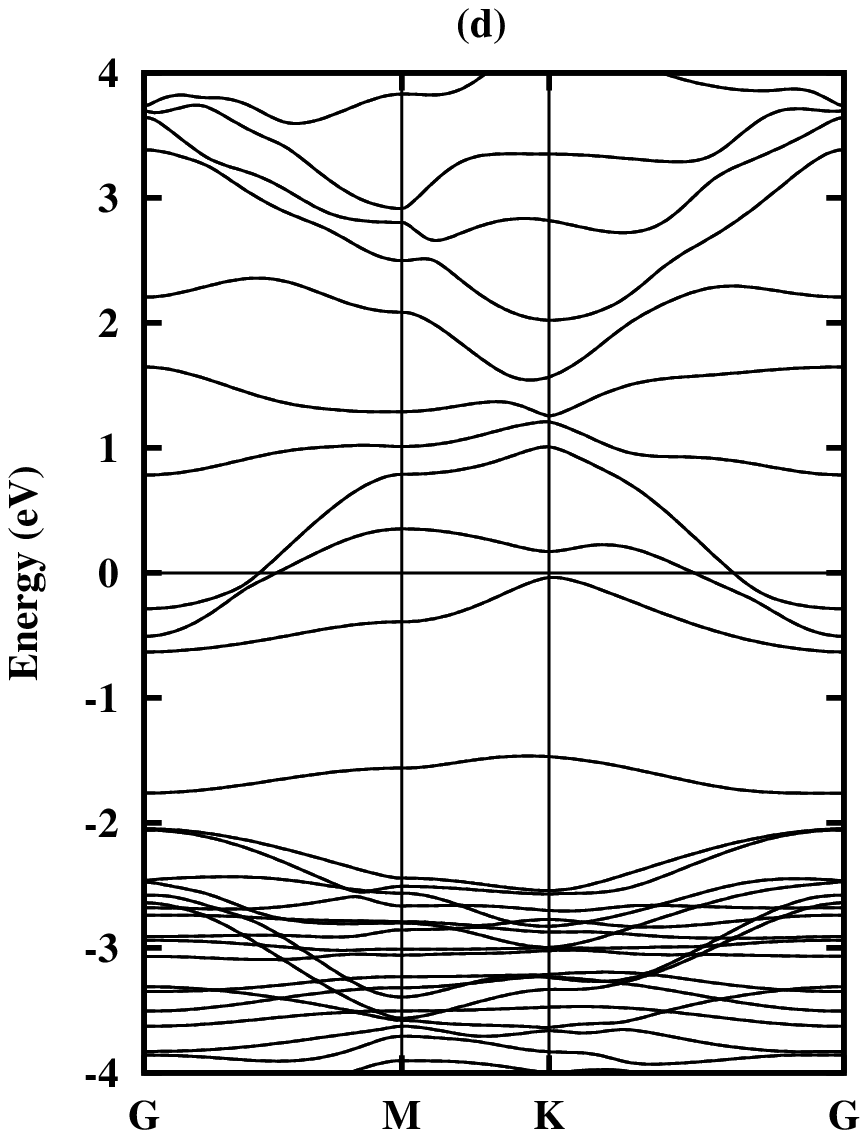}
\includegraphics[width=0.1\textwidth]{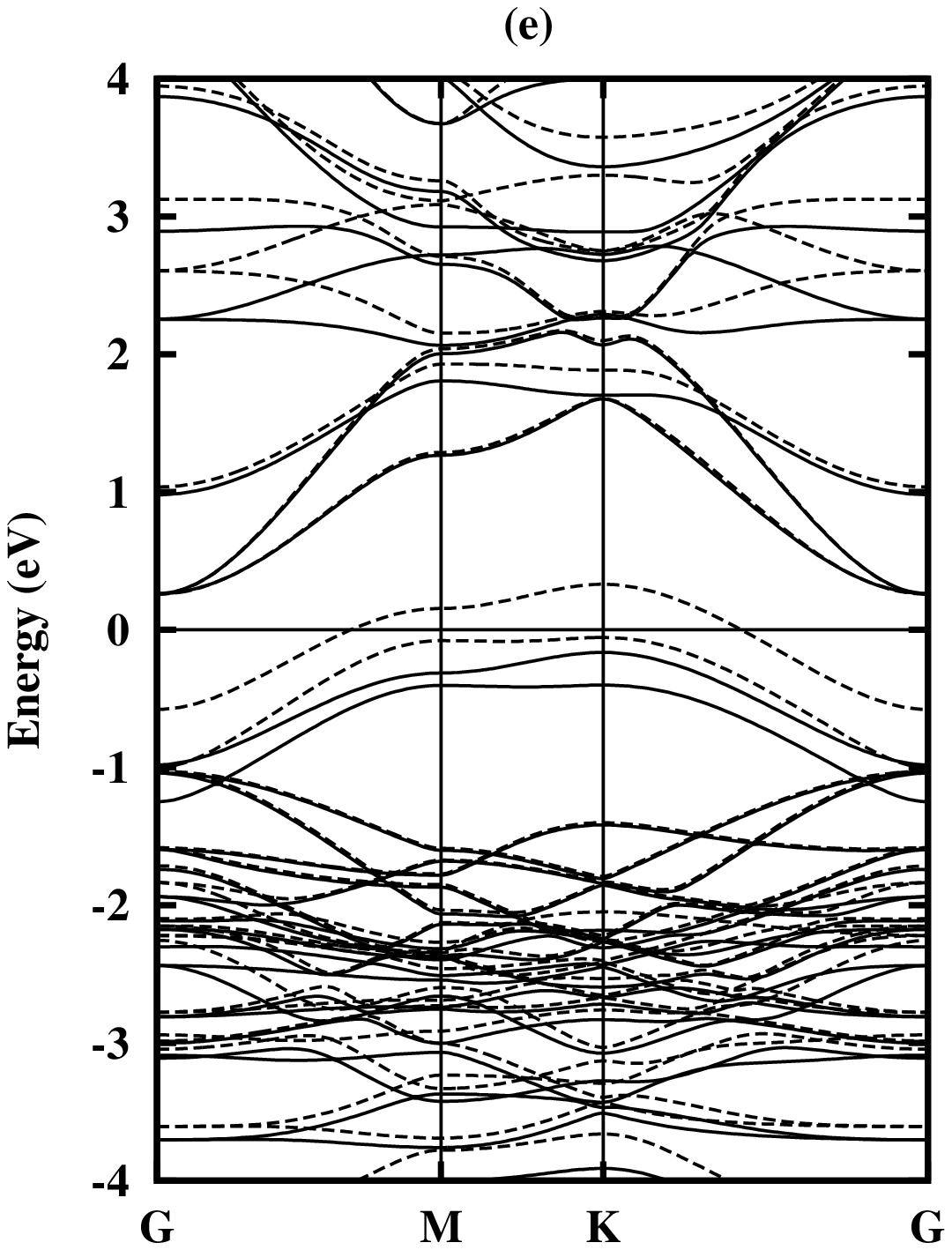}
\includegraphics[width=0.1\textwidth]{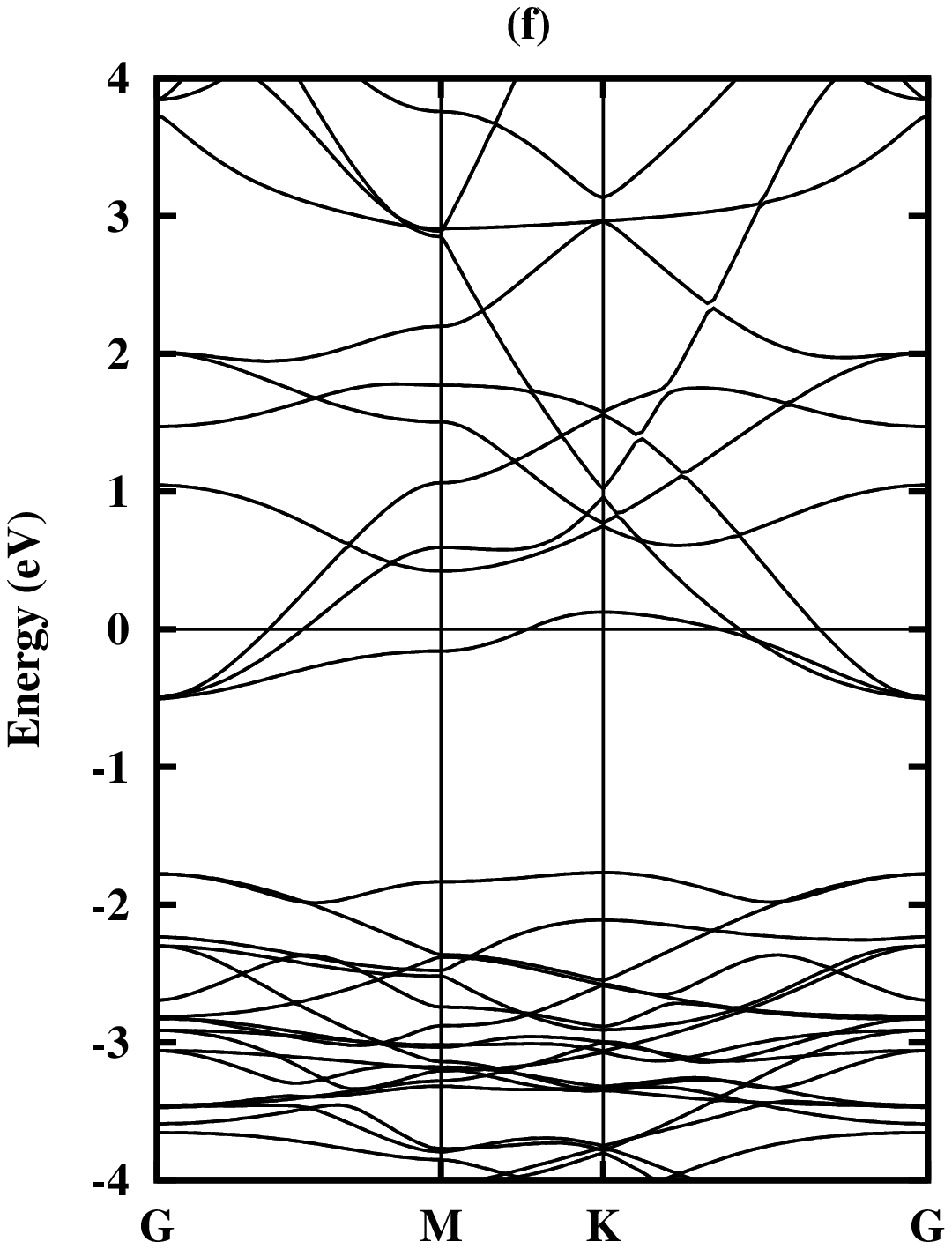}
\includegraphics[width=0.1\textwidth]{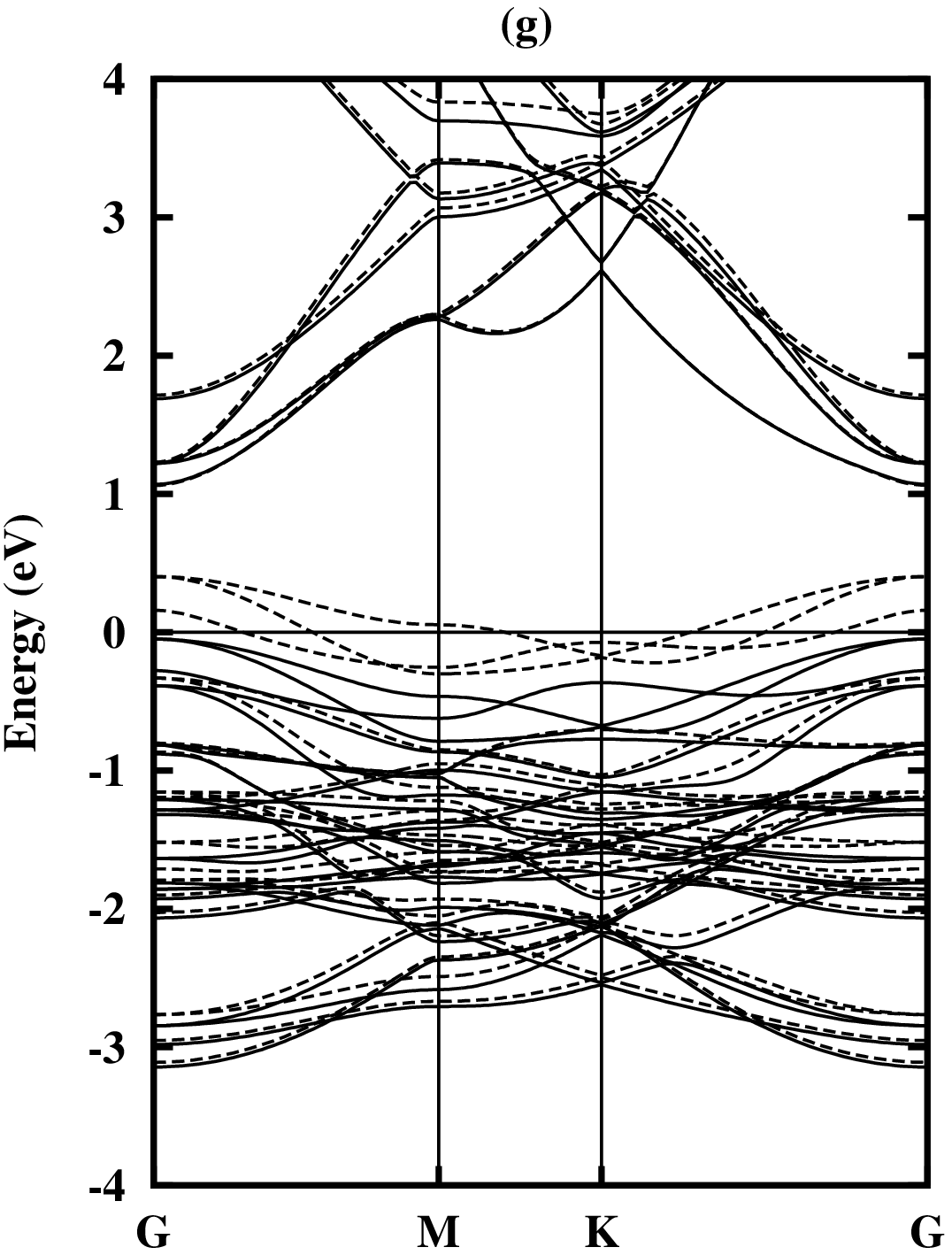}
\caption{The calculated band structure of pristine $2\times 2\times 2$ supercell of Li$_{3}$N (a), Li-I vacancy (b), Li-II vacancy (c), C-doped at Li-I (d), C-doped at Li-II (e), N-Vacancy (f), and C doped at N site in  $2\times 2\times 2$ supercell of Li$_{3}$N (g). The solid and dashed lines show the spin-up and spin-down electrons, respectively. The horizontal line shows the Fermi energy which is set to zero eV.}
\label{bands}
\end{figure}

Using the experimentally determined $c/a$ (1.062), we optimized the lattice constant $a$ of the $ \alpha $-Li$ _{3} $N with LSDA and found its value $\sim$ 3.42 {\AA}, whereas previously determined LDA and experimental values are 3.5 {\AA}\cite{A.C} and 3.648 {\AA},\cite{expRT,Hug} respectively.  
Whereas the GGA calculated value is 3.65 {\AA}, which is comparable with the previous GGA data.\cite{Shun2009}
Using the optimized lattice constants, the calculated $\Delta H_{f}$ is -3.10 (-1.69) eV with LSDA (GGA) whereas the experimental and the previous GGA values are -1.73 eV\cite{Sangster} and -1.59 eV,\cite{Shun2009} respectively. We repeated the same calculations using the QE code\cite{QE} with LSDA and got similar results to SIESTA. 
Hence, we conclude that LSDA underestimates the structural properties of bulk Li$_{3}$N. However, we didn't find any significant difference in the magnetic properties (which is the core of the subject) either using LSDA or GGA. We also calculated the electronic band structure of bulk  Li$_{3}$N, using both the non-spinpolarized (non-magnetic NM) and the spin-polarized (magnetic M) DFT calculations and  obtained similar results in both cases.\cite{GR2017} Our calculations predict a direct band-gap ($ \Gamma $-$ \Gamma $)  $\sim $ 2.0 eV, whereas its experimental value is $2.2$ eV. Both the values are in close agreement. {Note that band-gap was calculated as $|E_{c}-E_{v}|$, where $E_{c}$ ($E_{v}$) is the conduction (valence) band minimum (maximum) energy.} Bulk Li$ _{3} $N possesses an ionic nature  with maximum charge concentrated around the N sites and a slight charge distribution observed near the Li sites.\cite{GR2017} The calculated spin-polarized total and atom-projected (P) density of states (DOS),(see Ref.\onlinecite{GR2017}), reveal the absence of exchange splitting, which is a sign of non-magnetic behavior.  These results confirm that bulk Li$ _{3} $N has a non-magnetic and semiconducting character in its pure form which is consistent with the previous results.\cite{G.K} Therefore, here we propose that magnetism in bulk Li$ _{3} $N can be induced either by generating native defects (i.e., Li or N vacancy) or doping C at Li or N lattice sites. In this regard we analysed different defect-induced systems of Li$ _{3} $N by studying their electronic and magnetic properties in detail.  

\subsection{Li vacancy}

\begin{table}
\caption{Calculated formation energies (in eV) under Li-rich and N-rich conditions of different systems (Sys). The last column lists magnetic moments (M) per supercell calculated in units of $\mu_\mathrm{B}$. Values in parentheses are calculated with GGA.}
\begin{center}
\begin{tabular}{ccccccccc}
\toprule
\colrule
  Sys &    &  Li-rich & &   N-rich&&M\\
\hline
V$_{\rm {Li-I}}$&&    3.38 (2.54) &&    0.28 (0.84) &&1.00 (1.00)\\
V$_{\rm {Li-II}}$&&   1.72 (1.07)&&   -1.38 (-0.63)&&1.00 (1.00)\\
C$ _{\rm {Li-I}}$&&   3.06 (2.05) &&   -0.04 (0.35)&&0.00 (0.00)\\
 C$_{\rm {Li-II}}$&&  6.53 (5.73)&&    3.42 (4.03)&&1.00 (1.00)\\
V$_{\rm {N}}$&&       2.10  (2.54)&&    5.20  (4.24)&&  0.00 (0.00)\\
C$_{\rm {N}}$&&       2.41 (3.07) &&    5.51  (4.77)&&1.00 (1.00)\\
\hline
\end{tabular}
\label{F-energy}
\end{center}
\end{table}

To study defects-driven magnetism, we considered a $ 2\times 2\times 2 $ supercell of bulk Li$ _{3} $N and calculated its total energies in the NM and M states, and we confirmed that bulk pristine $ 2\times 2\times 2 $ supercell is NM. As  bulk Li$ _{3} $N has two types of Li-atoms, so we also analysed the band structure (see Fig.~\ref{bands}(a)) and electronic density of states which are shown in Fig.~\ref{pdos}(a). Fig.~\ref{bands}(a) clearly shows that $2\times 2\times 2 $ bulk Li$ _{3}$N is a semiconductor and the electronic dispersion increases along the $G-M$ direction. Fig.~\ref{pdos}(a) further shows that the valence band is mostly dominated by the N-$p$ orbitals whereas the conduction band has Li-$s$ character. The Li-II forms a narrow band as compared with Li-I atom and as Li-I-N bond length is smaller than the Li-II-N bond length, the hybridization of Li-I with N is stronger around 1 eV below the Fermi energy. The different behavior of Li in bulk Li$ _{3}$N suggests that defects can have different properties at Li-I and Li-II sites.

To search for possible magnetism induced by defects, we considered two kinds of Li vacancies in a $ 2\times 2\times 2$ supercell of bulk Li$ _{3} $N. We, therefore, propose two different systems, i.e., Li-I vacancy (V$ _{\rm Li-I} $) system and Li-II vacancy (V$ _{\rm Li-II} $) system (see Fig.\ref{str}).
The total energies of both the systems were calculated in the NM and M states, and we found that M state is more stable than the NM state. We, then, used the total energies and electronic structures in the M state for further analysis. 
We calculated the formation energy of each defect-induced system using Eq.~(\ref{vac}) under Li-rich and N-rich conditions, and the values are summarized in Table~\ref{F-energy}. The calculated formation energies for V$ _{\rm Li-I} $ and  V$ _{\rm Li-II} $ systems are 3.38 (0.28) eV and 1.72(- 1.38) eV in Li-rich (N-rich) conditions, respectively, thus it easier to generate V$ _{\rm Li-II} $ (in Li$ _{2} $N plane) than V$ _{\rm Li-I} $ under N-rich condition--similar results were also observed in the previous work~\cite{Shun2009}. Note that GGA values are
2.54 (0.84) eV and 1.07(- 0.63) eV in Li-rich (N-rich) conditions for V$ _{\rm Li-I} $ and V$ _{\rm Li-II} $ systems, respectively. Hence V$ _{\rm Li-II}$ is more stable under N-rich condition either using LSDA or GGA, which is also in agreement with the available experimental data.\cite{expRT,expRT2}  
The structural relaxation shows that the optimized bond lengths  Li$ _{\rm I} $-N  and Li$ _{\rm II} $-N in V$ _{\rm Li-I} $ system are 1.80 {\AA} and 1.97 {\AA}, respectively which are comparable with those of pristine Li$ _{3} $N which are 1.81 {\AA} and 1.97 {\AA} respectively. Thus, no significant structural changes are noticed in V$ _{\rm Li-I} $ system. On the other side, generating a Li-II vacancy causes the Li$ _{\rm II} $-N optimized bond length to reduce by 0.1 {\AA} when compared with that of pristine Li$ _{3} $N, while bond lengths remain unaffected along the $c$-direction.

The spin-polarized calculations (see Fig.\ref{bands}(b,c)) reveal that in both the systems (V$ _{\rm Li-I},\rm V _{\rm Li-II}$) Li vacancies induce magnetic moments of 1.0 $\mu_{\rm{B}}$, and the magnetic moments are mostly localized around the N atoms near to the Li vacancies. 
Fig.\ref{bands}(b) shows the spin-polarized band structure of V$ _{\rm Li-I}$, where one can see that the spin-up states are occupied and the spin-down states are partially occupied. Large spin-polarization near the Fermi energy in the valence band indicates that Li vacancy induces holes in Li$_{3}$N. It is expected that creating a Li vacancy (Li-I or Li-II) can cause the oxidation state of N-anion to change from N$ ^{3-} $ (pristine Li$ _{3} $N) to N$ ^{2-} $ (V$ _{\rm Li-I} $ or V$ _{\rm Li-II} $ system), thus inducing a hole in the minority spin states of N-$p$ orbitals, consistent with the Hund's rule. A local magnetic moment of $\sim$ 0.47 $\mu_{\rm B}$ is observed at N  site close to the Li vacancy in V$ _{\rm Li-I} $ system and $\sim$ 0.39 $\mu_{\rm B}$ for V$ _{\rm Li-II} $ system. The band structure of V$ _{\rm Li-II}$ (Fig.\ref{bands}(c)) has a similar feature to V$ _{\rm Li-I}$ except large minority bands at the Fermi energy.  
To elucidate the atomic origin of Li-vacancy driven magnetism in Li$_{3}$N, we show the calculated spin-polarized total and atomic projected partial density of states in Fig.~\ref{pdos}(b,c).     
A narrow minority band in the bandgap of Li$ _{3} $N, including the Fermi energy ($E_{\rm F}$), is evident from  the total DOS and PDOS plots of V$ _{\rm Li-I} $ system(Fig.~\ref{pdos}(b)), which is contributed mainly by the $p$-orbitals N atoms nearest to the Li-I vacancy. One can see that the spin-up states are completely occupied, whereas the spin-down states are partially occupied consistent with the band structure. The PDOS shows that the partially unoccupied spin-down states are mainly contributed by the N-$p$ orbitals. Note that the N-$p$ orbitals are completely occupied in the pristine  Li$_{3}$N. Hence, Li vacancy introduces holes which are mostly localized around the N atoms, and it is expected that these holes will mediate the long-range (ferro)magnetism in Li$ _{3}$N.
A small spin polarization is also seen in the distant N-$p$ orbitals. No spin split is obvious at the Li sites (Li-I and Li-II). In Fig.\ref{pdos}(b), Li-II (Li-I) are  near (far) to V$ _{\rm Li-I}$. The only difference observed for Li-$s$ orbitals, when compared with those of pristine Li$ _{3} $N [Fig.~\ref{pdos}(a)] is that these orbitals are driven close to the $E_{\rm F}$ after inducing  the (Li-I) vacancy, thus reducing the bandgap. Hybridization of Li-$s$ and N-$p$ orbitals can be observed at $\sim$ 2 eV below the Fermi energy level. 
We also observed a slight different behavior in the PDOS of V$_{\rm Li-II} $ system(Fig.~\ref{pdos}(c)), where one can also see some small spin-polarization at the Li sites (mainly Li-II, which is near to V$ _{\rm Li-II}$), and the Li-$s$ orbitals are shifted more towards the Fermi energy as compared with V$ _{\rm Li-I}$ system(Fig.~\ref{pdos}(b)). One can also see small exchange splitting, which leads to magnetic moment, at N site in V$ _{\rm Li-II}$.
To further see the spin-polarization induced by Li vacancy, we show the spin density around the Li-II vacancy, e.g., in Fig.\ref{sdos}(a), where one can clearly see a large spin-polarization around the N atoms near to Li-vacancy. Similar magnetic moments were also observed in the GGA calculations.

\subsection{C doped at Li sites}
\begin{figure}[]
\includegraphics[width=0.1\textwidth]{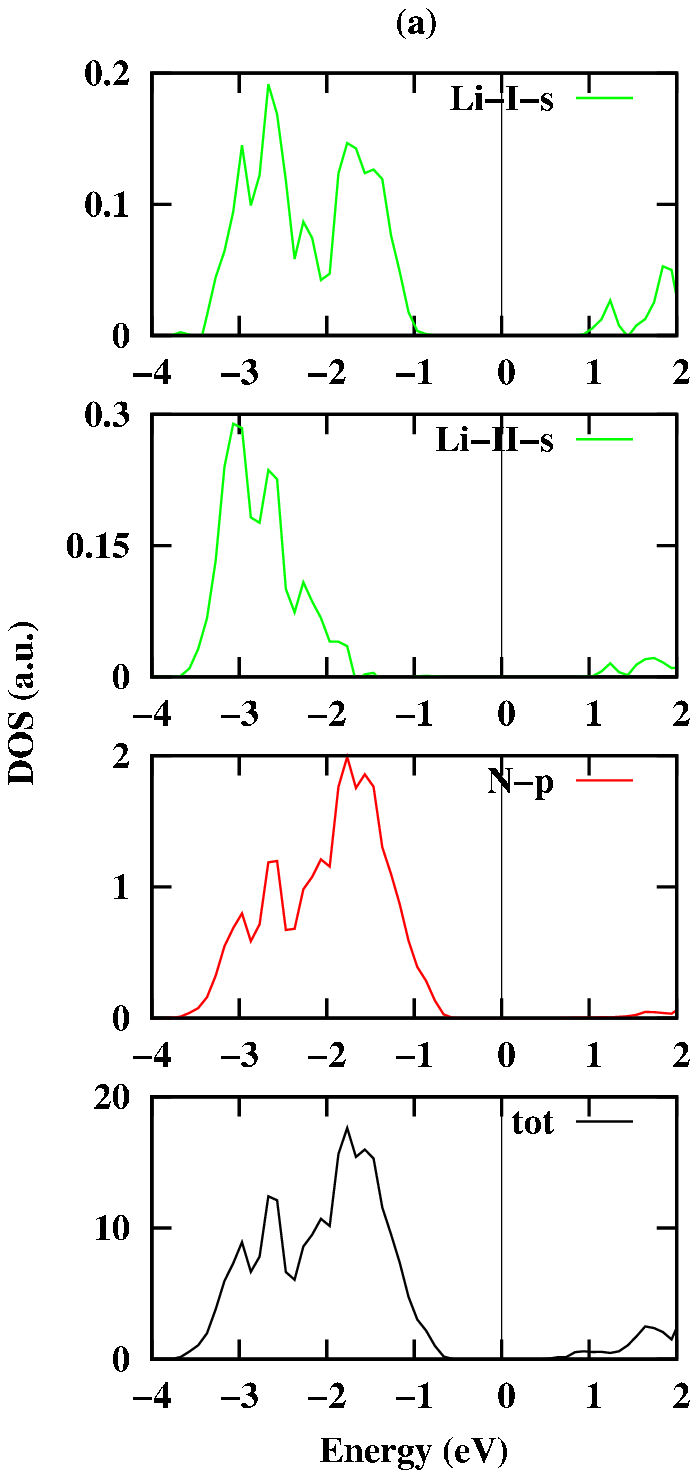}  
\includegraphics[width=0.1\textwidth]{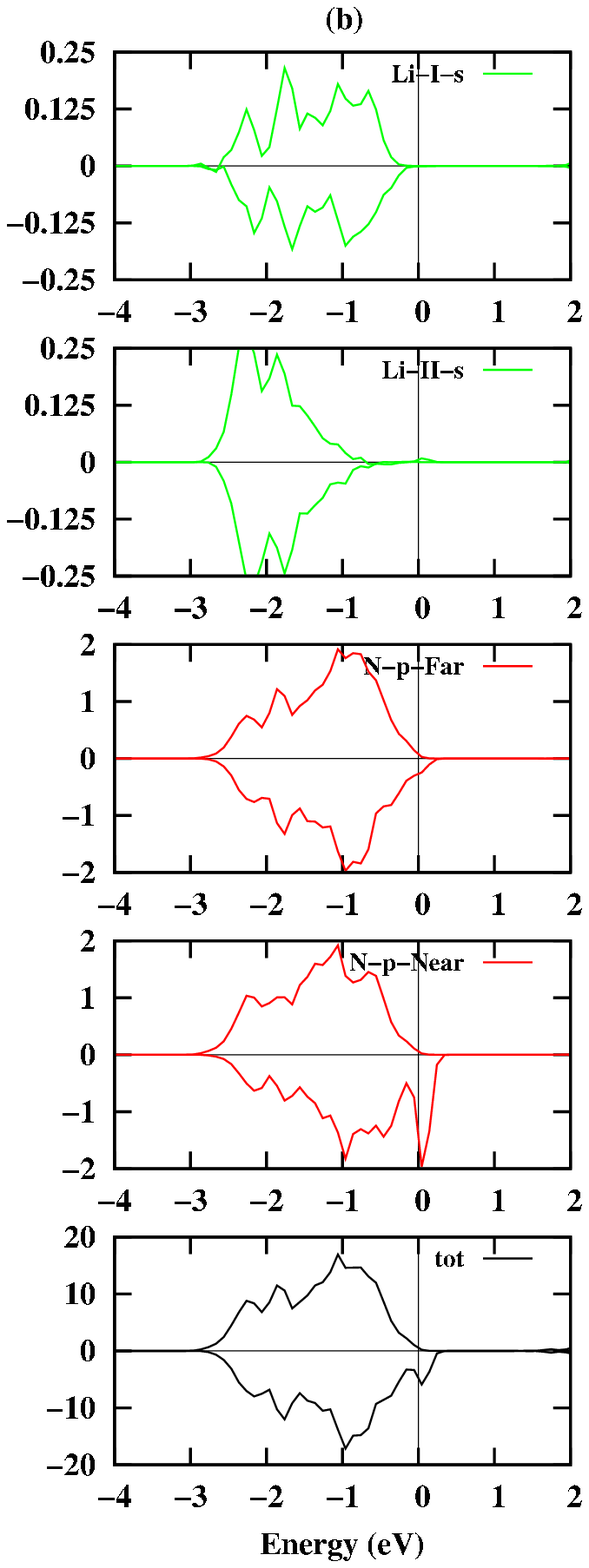}  
\includegraphics[width=0.1\textwidth]{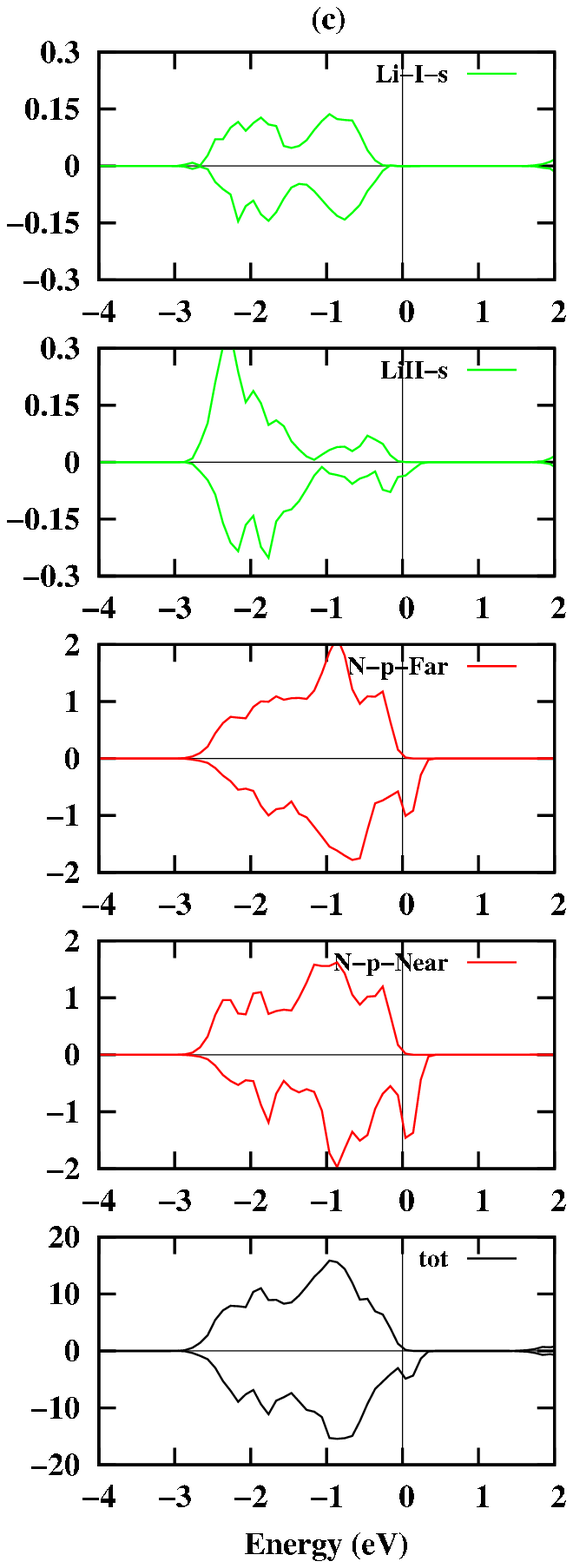}  
\includegraphics[width=0.1\textwidth]{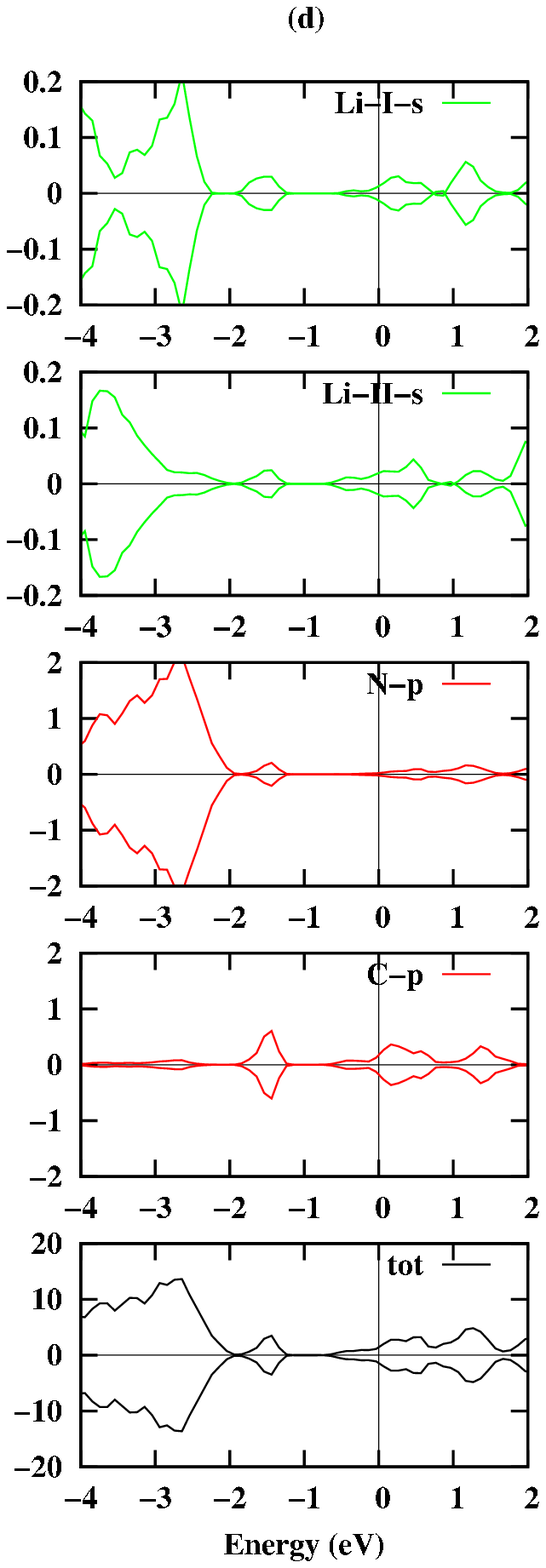}\\
\includegraphics[width=0.1\textwidth]{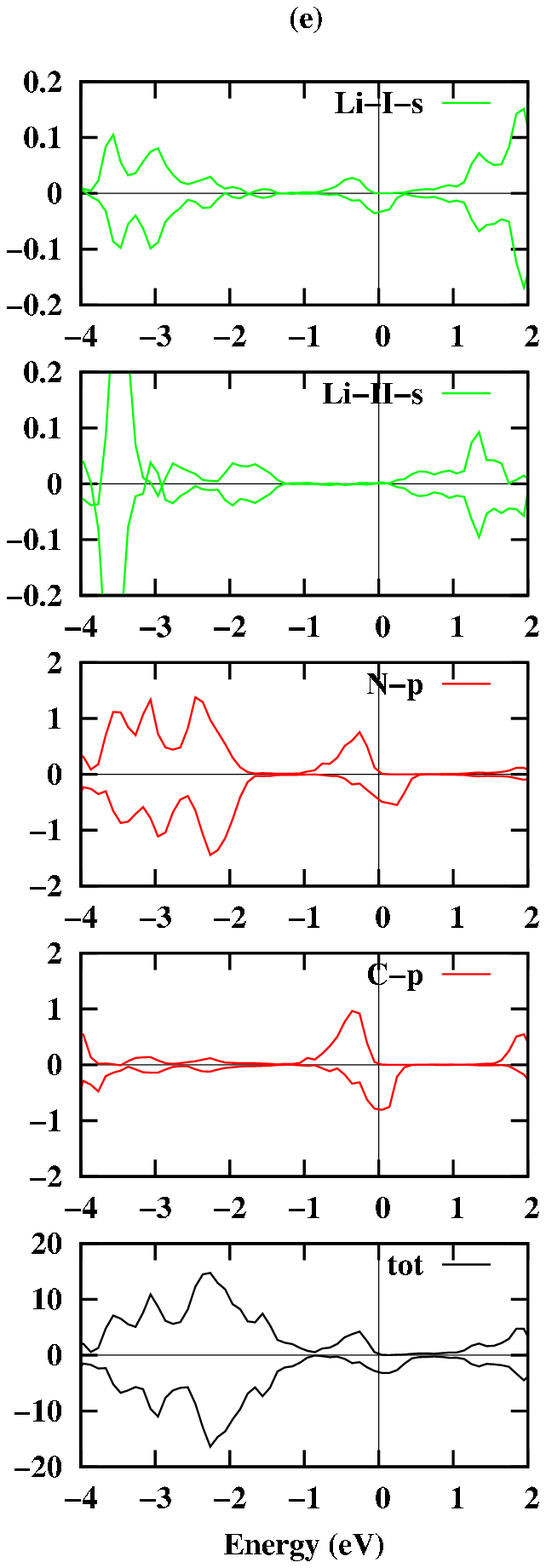}  
\includegraphics[width=0.1\textwidth]{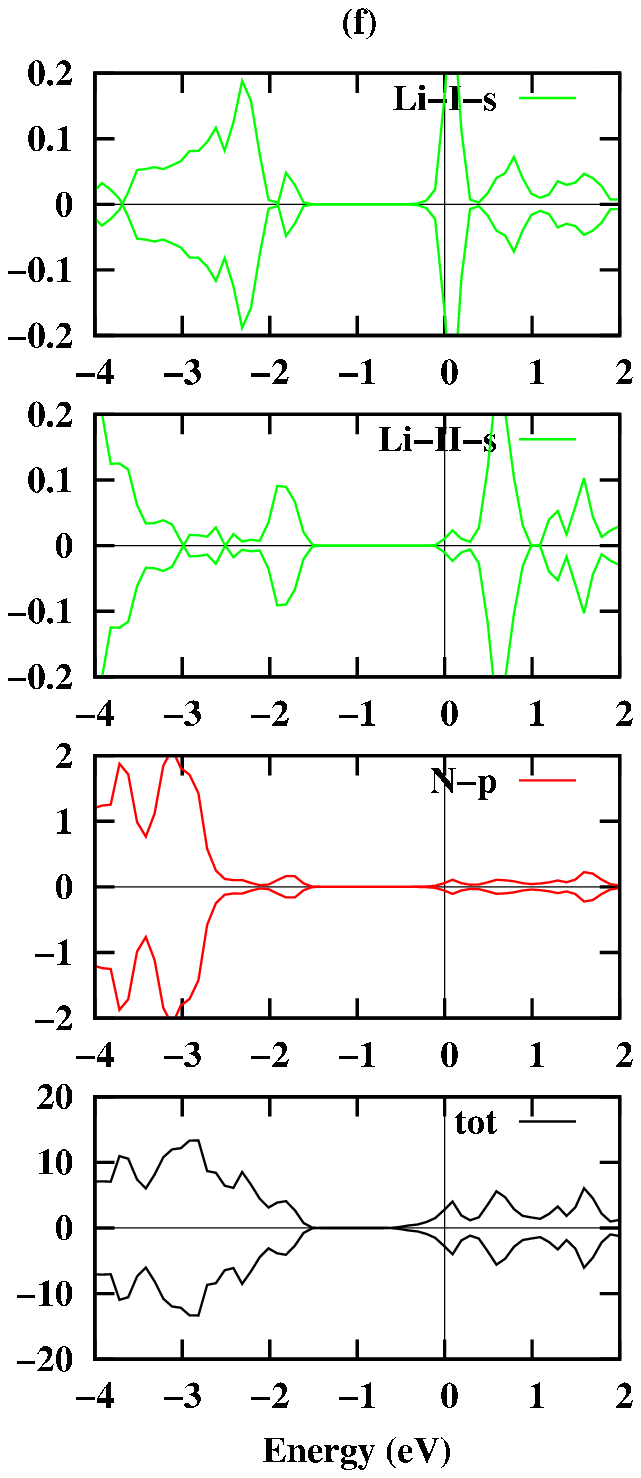}  
\includegraphics[width=0.1\textwidth]{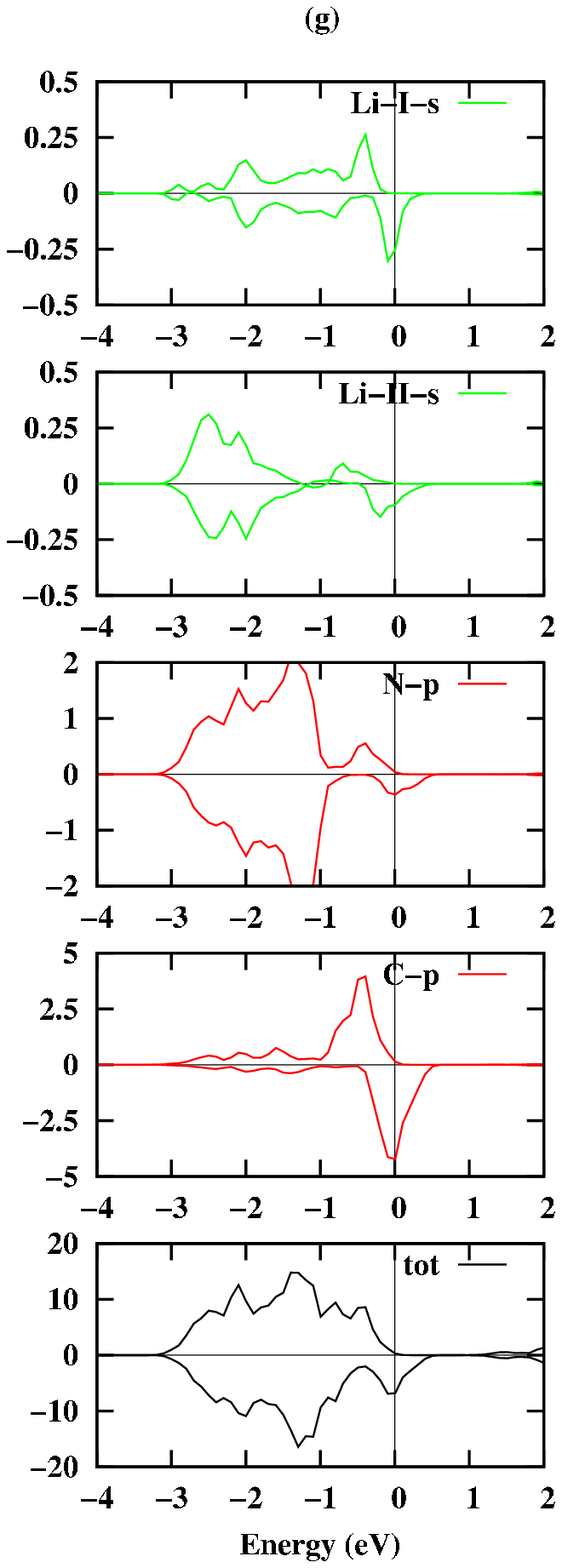}  
\caption{ (Color online) The calculated spin-polarized
electronic total and atom projected density of states of pristine $2\times 2\times 2$ supercell of Li$_{3}$N (a) Li-I vacancy (b), Li-II vacancy (c), C doped at Li-I (d), C doped at Li-II (e), N-vacancy (f), and C doped at N site in Li$_{3}$N(g). Black lines show the total DOS, whereas red and green lines represent the $p$ and $s$ orbitals of N/C and Li atoms, respectively. The vertical lines show the Fermi energy which is set to zero eV.  }
\label{pdos}
\end{figure}

In the above sections, we have shown that V$ _{\rm Li-I} $ and V$ _{\rm Li-II} $ can induce magnetism due to holes which reside on N atoms. It would also be interesting to see whether carbon can induce magnetism in Li$ _{3} $N when doped either at Li-I or Li-II site. Previous work show that TM (Fe, Co, Ni, Cu) prefer the Li-I site and can induce a large magnetization in Li$_{3}$N.~\cite{wu2011,wu2009}
Here we investigate doping of C atom at the two possible Li (Li-I and Li-II) sites in bulk Li$ _{3} $N and propose two further systems; C doped at Li-I site (C$ _{\rm Li-I} $) system and C doped at Li-II site (C$ _{\rm Li-II} $) system. The calculated formation energies (using Eq.~\ref{C-doped}) for C$ _{\rm Li-I} $ and C$ _{\rm Li-II} $ systems are 3.06 (-0.04) eV and 6.53 (3.42) eV under Li-rich (Li-poor) condition, respectively. Therefore, it is expected that doping C at Li-I site( C$ _{\rm Li-I}$) is thermodynamically more stable under N-rich condition as compared with C$ _{\rm Li-II}$, similar to TM in Li$ _{3} $N.~\cite{wu2011} Such thermodynamic stability is accompanied by a structural distortion around the C atoms.
We also analysed the atomic positions, and 
the structural analysis of the C$ _{\rm Li-I} $ system reveals that the optimized C-N bond length is 1.44 {\AA} which is 1.81 {\AA} for Li$ _{\rm I} $-N in pristine Li$ _{3} $N. Thus replacing Li-I by C  atom reduces the bond length by $\sim$0.40 {\AA}.
Similarly when the Li$ _{\rm II} $-N and Li$ _{\rm II} $-Li$ _{\rm II} $ bond lengths in C$ _{\rm Li-I} $ system are compared with those of pristine Li$ _{3} $N, we found the mentioned bond lengths are elongated by $\sim$ 0.1 {\AA} and 0.07 {\AA}, respectively. Thus significant structural changes are observed for C$ _{\rm Li-I} $ system, which is attributed to the smaller atomic size of C as compared with the atomic size of Li. 

To search for possible magnetism in C$ _{\rm Li-I} $ system, we calculated the band structure using spin-polarized DFT. The spin-polarized electronic band structure (see Fig.~\ref{bands}(d)) shows no spontaneous magnetism (no exchange splitting) and the calculated magnetic moment is zero. Though C at Li-I site is thermodynamically stable as compared with Li-II, but does not induce any magnetism.  Fig.~\ref{bands}(d) clearly shows that doping C at Li-I induces metallicity in Li$ _{3} $N. Comparing Fig.~\ref{bands}(d) with Fig.~\ref{bands}(a), one can see that the Fermi energy lies in the conduction band, and there are three bands that cross the Fermi energy. Our detailed PDOS (see Fig.~\ref{pdos}(d)) analysis show that these bands are mainly contributed by the C and N $p$-orbitals. One can also see some C-driven Li-$s$ states at the Fermi energy.  
Previous work indicate that Co or Ni doping at Li-I site can
reduce the energy band gap and even change Li$ _{3} $N from a
semiconductor to a metallic-like conductor, which has the
advantage of both electronic and ionic conduction.\cite{wu2011,wu2009} 
As Li$ _{3} $N shows negligible electronic conduction,\cite{nazri} which restricts its application as anode
material in rechargeable lithium batteries. Therefore,
C doping at Li-I site changes Li$_{3}$N from a semiconductor to a metallic-like conductor and it can help in the electronic conduction that may have applications in rechargeable lithium batteries.
Doping C at Li-I site was expected to induce magnetism in bulk Li$_{3}$N but the result was contrary to our expectation due to structural distortion (and small C-N bond length) which is quenching the spin-magnetic moments of $p$-orbitals of C atoms. Due to a small C-N bond length, the $p$ orbitals overlap so densely, thus leaving no uncompensated spin  which is further confirmed by the total and atom projected DOS of C$ _{\rm Li-I} $ system. The absence of magnetism in this case is different from TM doped Li$ _{3} $N where TM carries local spin magnetic moment.\cite{wu2011,wu2009}
\begin{figure}[]
\includegraphics[width=0.4\textwidth]{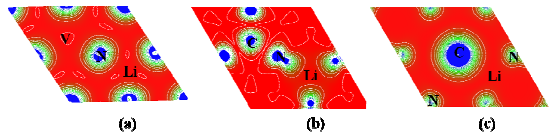}  
\caption{ (Color online) Spin density of V$_{\rm Li-II}$ (a), C$_{\rm Li-II}$ (b), and C$_{\rm N}$ (c), where atoms are labeled as C, N, Li, and the Li vacancy is represented by V.}
\label{sdos}
\end{figure}

Contrary to C$ _{\rm Li-I} $, doping C at Li-II site results in less structural changes. The calculated C-N bond length is $\sim$ 2.04 {\AA} which was 1.97 {\AA} (Li$ _{\rm II} $-N) in pristine Li$ _{3} $N before replacing Li-II by C, thus C-N bond length is increased by 0.07 {\AA} after doping. Similarly, the bond lengths C-Li$ _{\rm II} $ (C$ _{\rm Li-II} $ system) and Li$ _{\rm II} $-Li$ _{\rm II} $ (pristine Li$ _{3} $N) are 1.92 {\AA} and 1.97 {\AA}, respectively. Whereas no significant changes in bond lengths are noticed along the $c$-direction. 
As the relaxed bond length of C-N (2.04\AA) in C$ _{\rm Li-II} $ is larger than the  C-N bond length (1.44\AA) in C$ _{\rm Li-I}$ and less structural distortion is observed in C$ _{\rm Li-II}$, hence weak hybridization, which will lead to some magnetic moments, is expected.
Doping C at Li-II site creates a hole in the N orbital and therefore  a net magnetic moment of 1.0 $\mu_{\rm B}$ per cell is observed. Detailed analysis of Mulliken orbital populations demonstrates that the local magnetic moment observed for C is $\sim$ 0.316 $\mu_{\rm B}$, and each three N atoms surrounding the C atom have a local magnetic moment $\sim$ 0.23$\mu_{\rm B}$. This is interesting to note that the magnetic moment is not mostly localized around the C atoms, but also spreads to its nearest N atoms. 
The spin-polarized electronic band structure of C$ _{\rm Li-II} $ system is also shown in Fig.~\ref{bands}(e). The spin-up states are completely occupied and has 
band gap ($\ge 1.0\,$eV) at $G$-point, on the other hand the spin-down band is partially occupied and has large states at the Fermi energy\,--- a half-metallic band structure. 
A further insight into the magnetic nature of the C$ _{\rm Li-II} $ system is gathered from the total and atom-projected DOS (see Fig.~\ref{pdos}(e)), which displays an impurity-driven band in the band gap of the Li$ _{3} $N that includes $ E_{\rm F} $. This impurity band is mainly contributed by the partially occupied minority spin states of C and N $p$  orbital. Hybridization between the N-$p$ and C-$p$ orbitals can be viewed particularly near $E_{\rm F}$ where the majority spin states are completely filled while the minority spin states are partially empty, thus leading to a significant spin-split at the Fermi energy. A small spin polarization at the Li-I site is also visible in the PDOS plot. The hybridization of Li-$s$ orbitals with the $p$-orbitals of N and C near the Fermi energy and also deep in the valence band is visible. The impurity bands contributed by the C atoms have a high density in the majority spin states near the top of the valence band. C doping has also driven the N-$p$ states closed to the Fermi level, where its minority spin states have been moved higher above the valence band maximum as compared to the C-$p$ minority  spin states. Fig.\ref{sdos}(b) further illustrates that the spin density is not only localized at the C site, but also spread over the nearest N atoms. The spin density around the N atoms are not spherically distributed but has dumbbell like shape where the density decreases towards the Li atom.

It is important to discuss that C$ _{\rm Li-I}$ is more stable and non-magnetic either using LSDA or GGA calculations. The Li-I type atoms connect the Li$ _{2} $N layers and the bond length of Li$_{\rm I}$-N is smaller than the Li$_{\rm II}$-N. Carbon also has a smaller atomic size as compared with Li atomic size, hence strong C-N bond is expected when C is doped at Li-I site.
The Li$ _{3} $N structure can be thought as  Li$_{2}$-IILi$_{1}$-IN.
The Li-I atoms form a free layer and it would be easy to dope Li-I atoms with the impurity atoms. On the other hand, the Li-II atoms form a network in the Li$ _{2} $N plane in the Li$ _{3} $N structure, and hence it is expected that the N-$p_{x}/p_{y}$ orbitals will overlap with the Li atoms and it can cost more energy to  dope impurity atoms at Li-II site. The minority unoccupied spin of N/C-$p$ orbitals are responsible for the larger formation energy. It is also important to compare the magnetism driven by C in Li$ _{3} $N with TM doped in Li$ _{3} $N. TM doped at Li-I site is thermodynamically stable and has magnetic ground state,\cite{wu2011,wu2009} in our case doping C at   
Li-I site is thermodynamically more stable as compared with Li-II site. However, C does not induce magnetism when doped at Li-I site, but it only induces magnetism when doped at Li-II site. The TM carries a local $d$-electron magnetic moment, so with in the diluted limit magnetism can be expected in TM-doped at either sites (Li-I or Li-II) in Li$ _{3} $N.  
\subsection{N vacancy}
Another possible intrinsic defect in Li$ _{3} $N is N-vacancy and in this section we focus on the electronic structure of N-vacancy V$ _{\rm N} $, which is created at the center of Li$ _{2} $N plane. The calculated defect formation energy (using Eq.~\ref{vac}) is 2.10 (5.20) eV in Li-rich (N-rich) environment. N vacancies in Li$ _{3} $N are possible in Li-rich condition. The defect formation energy of V$ _{\rm N} $ is larger than the Li-vacancies, and hence it indicates that the most probable defects in Li$ _{3} $N could be the Li vacancies either Li-I or Li-II.
The Li$ _{\rm I} $-N and Li$ _{\rm II} $-N bond lengths are found
to be $\sim1.76$\AA\, and $\sim 1.92$\AA, respectively. Similarly the calculated Li$ _{\rm II} $-Li$ _{\rm II} $ bond length in V$ _{\rm N} $ system is 2.08\AA, whereas in pure system it is 1.97\AA.
When compared with the pure system, the optimized bond lengths Li$ _{\rm I} $-N and Li$ _{\rm II} $-N of V$ _{\rm N} $ system are reduced by 0.05 {\AA}, while Li$ _{\rm II} $-Li$ _{\rm II} $ is increased by 0.11 {\AA} after relaxing the structure.  Each N in Li$ _{3} $N is surrounded by eight Li atoms and removing a single  N atom results in the formation of a local Li$_{8}$-type cluster, which is non-magnetic. Therefore, no magnetism is induced through N vacancy in Li$ _{3} $N. 
The spin-polarized electronic band structure (Fig.~\ref{bands}(f)) reveals no signature of magnetism and zero magnetic moment is observed. Fig.~\ref{bands}(f) shows that the Fermi energy is shifted towards the conduction band as compared with Fig.~\ref{bands}(a), hence V$ _{\rm N} $ behaves as $n$-type impurity in Li$ _{3} $N and large N vacancies can result metallicity in Li$ _{3} $N. To further analyse the origin of impurity-driven bands in the band gap, we show the calculated total and atom-projected DOS of the V$ _{\rm N} $ in Fig~\ref{pdos}(f). Fig~\ref{pdos}(f) shows that the impurity band in Fig.~\ref{bands}(f) is mainly contributed by the Li-$s$ electrons, particularly by Li-I . Note that the Li-$s$ and N-$p$ orbitals were completely occupied in the pristine Li$ _{3} $N, but the N vacancy delocalises these orbitals and can give metallic-like character in Li$ _{3} $N.  Absence of the spin split in the DOS emphasizes that creating a N vacancy in the bulk-Li$_{3}$N does not induce magnetism. Similar results were also observed in the GGA calculations.

\subsection{C doped at N site}
To dig another possible source of magnetism in bulk Li$ _{3} $N, we considered doping C at the N site (C$ _{\rm N} $) and investigated its electronic and magnetic structure. The calculated defect formation energies of C$ _{\rm N}$ under N-rich and Li-rich conditions (using Eq.~\ref{C-doped}) are 5.51 eV and 2.41 eV, respectively, which is less than C$ _{\rm Li-II} $ system. Thus C doping at the N site is thermodynamically more favourable as compared with C$ _{\rm Li-II}$. When the relaxed atomic coordinates of this system are analysed, it is noticed that the optimized bond lengths Li$ _{\rm II} $-C and Li$ _{\rm II} $-Li$ _{\rm II} $ are increased by 0.06 {\AA}, and the Li$ _{\rm I} $-C is increased by 0.07 {\AA}, when C is doped at the N site in bulk Li$ _{3} $N. To further confirm the observed bond length expansion of the C$ _{\rm N} $ system, we performed extra calculations and optimized the lattice constant $a$ of $ 2\times 2\times 2$ supercell of the C$ _{\rm N} $ system using the total energy calculation method. The calculated optimized lattice constant $a$ of C$_{\rm N}$ system is 3.47{\AA}, and that of pristine system is 3.42{\AA}. The lattice constant of the unit cell of C$ _{\rm N} $ system is increased by 0.05 {\AA}. Hence, it is to infer that the Li$ _{3} $N crystal structure shows expansion after doping C at the N site. Such increase in the bond length is mainly attributed to larger atomic size of C as compared with the atomic size of N.
Our spin-polarized calculations show that C induces a large magnetic moment when doped at N site in Li$_{3}$N---the calculated total magnetic moment with GGA and LSDA is 1.0 $\mu_{\rm B}$, which is arising mainly due to the uncompensated spin in the C-$p$ orbitals. In C$ _{\rm N} $ system C is doped at anion site (N), hence it is expected that C will be in C$ ^{3-} $ anioin ($s^{2}p^{5}$) state. Detailed analysis of the Mulliken orbital populations reveals that the maximum contribution of $\sim$ 0.82 $\mu_{\rm B}$ is received from the C atom, and some magnetic moment of $\sim$ 0.09 $\mu_{\rm B}$ is induced in the surrounding N atoms. The far N atom also has a local magnetic moment of $\sim$ 0.02 $\mu_{\rm B}$. The strength of local magnetic moments is found to decrease with increasing distance from the C atom. Nonzero spin-polarization of the Li orbitals was also observed.  

Figure~\ref{bands}(g)shows the spin-polarized band structure of C$ _{\rm N} $ where large spin-polarization near the Fermi energy is visible. The Fermi level lies on the top of the valence band, similar to Li-II vacancy, and the majority spin states are completely occupied whereas the minority spin states are partially occupied--a half-metallic band structure. Similar to Li-II vacancy system, C mainly polarises the valence band suggesting that C introducing holes in Li$_{3}$N.
The total and atom-projected DOS calculations interpret further the magnetic behavior of the C$ _{\rm N} $ system (see Fig~\ref{pdos}(g)). An impurity derived peak contributed by the C-$p$ orbitals is observed in the band gap of Li$ _{3} $N, which includes $E_{\rm F}$. The PDOS also shows the hybridization of Li-$s$, N-$p$, and C-$p$ orbitals at $E_{\rm F}$. The Fermi energy region is dominated by the C-$p$ orbitals, having a high density of states. C doping  also induces spin polarization in N-$p$ and Li-$s$ orbitals. The majority and minority spin states of C are localized in $E_{\rm F}$ region, below and above the valence band maximum, respectively. C doping  also raises the Li-$s$ and N-$p$ orbitals in energy. The magnetism driven by C ($ s^{2}p^{5} $) doping at N ($ s^{2}p^{6} $) site is due to the hole in the C-$p$ minority spin state, which results in a total magnetic moment of 1.0 $\mu_{\rm B}$. We also analysed the spin-density of C$ _{\rm N}$ (see Fig.~\ref{sdos}(c)) and the magnetic moment is mainly localised around the C atom and small polarization at the N site can also be seen. Note that in C$ _{\rm Li-II}$ case the spin density was not mainly localised around the C atom, but spread over the N atoms as well.

In the above section, we found that C$ _{\rm N}$ is magnetic and has a smaller formation energy than C$ _{\rm Li-II}$. To further qualify C$ _{\rm N}$ system experimentally, it is very essential to investigate the magnetic coupling between the C spins, so we doped two C atoms at the N sites in a $2\times 2\times 2$ supercell.
The total energies were calculated in the ferromagnetic (FM) and antiferromagnetic (AFM) states,
to estimate the strength of exchange interaction $J$ which is predicted as, $ \Delta E = E_{AFM} - E_{FM} $. Here $E _{AFM} (E_{FM}) $ is the total energy of the two C-doped C$ _{\rm N} $ system per supercell in the AFM (FM) state.
We consider two different configurations of the two C-doped C$ _{\rm N} $ system, depending on the position of the C atoms, where one of the C atoms ({denoted as C$_{0}$ in Fig.\ref{str}(b)}) is fixed at the center while the position of the other C atoms ({denoted as C$_{1}$ and C$_{2}$ in Fig.\ref{str}(b)}) is altered. The spin polarized calculations confirm that the electronic band structures retain their half-metallic character for both the systems.
The magnetic moment, in both cases,  is determined to be 2.0 $\mu_{\rm B}$ per supercell, i.e.,1.0  $\mu_{\rm B}$ per C atom.
In one of the systems, we doped the second C (C$_{1}$) at the center of the adjacent Li$_{2}$N layer, such that both of the C atoms occupy the central positions in the adjacent Li$ _{2} $N layers, and the C atoms are connected through Li-I atoms, i.e, it forms a linear chain C-Li-I-C parallel to the $c$-axis, and C$_{0}$-C$_{1}$ distance is 3.63\AA. 
We found that FM state is more stable than AFM state by 0.12 eV/supercell, and the defect formation energy is 5.54 eV per C atom. Using the calculated $ \Delta E$, {the estimated $J$ within the Heisenberg nearest neighbor model is about 30 meV.\cite{deltae}}
In the other model, one of the C is fixed at the center of Li$ _{2} $N layer while other C atom (denoted as C$_{2}$) is positioned in the adjacent Li$ _{2} $N layer at a distance (C$_{0}$-C$_{2}$) of 4.98 {\AA} w.r.t the first C$_{0}$ atom. {In this case $\Delta E$ $\sim$ 0.07  eV/supercell, and the estimated $J$ is about 8.75 meV\cite{deltae}}, whereas the defect formation energy is 5.47 eV per C atom. 
Therefore, in both cases we always found that FM state is more stable than the AFM, and $J$ decreases with C-C separation. No clustering of C atoms can be expected. Therefore, based on our extensive DFT calculations, ferromagnetism is expected if C is engineered at the N site in Li$_{3}$N.

\section{Summary}
To summarize,  using \textit{ab-initio} calculations, we investigated possible defects-driven  magnetism in bulk $\alpha $-Li$ _{3} $N. We analysed that Li vacancies (Li-I, Li-II) are significant for establishing  magnetic moments ($ \sim 1.0$ $ \mu_{\rm B} $) and half-metallic character in Li$ _{3} $N. Li vacancy induces a  hole localized in the N-$ p $ orbitals which is expected to mediate long range (ferro)magnetism in Li$ _{3} $N.  The formation energy of Li-II vacancy under N rich condition is least among all kinds of vacancies, and therefore expected to be more probable defect in bulk Li$ _{3} $N.  N vacancy in Li$ _{3} $N is non-magnetic but can be crucial for conduction. Our calculations also revealed  that carbon doping at the atomic sites (Li-II, N) can  exhibit magnetism ($ \sim 1.0$ $ \mu_{\rm B} $) in Li$ _{3} $N, whereas C doping at Li-I site does not favor magnetism due to large structural distortion, but having low formation energy under N rich condition--this defect is significant for electronic conduction in Li$ _{3} $N. The calculated formation energies emphasized that C doping at N site is more favorable than at Li-II site. The impurity induced hole in C$ _{\rm N} $ system and is expected to mediate ferromagnetism in the host material,  and the spin density is found largely localized at the dopant site. We also examined the magnetic coupling between impurity induced spins in  C-doped Li$ _{3} $N and analysed that it has a ferromagnetic ground state. 
Our analysis suggests that defect-driven magnetism in Li$ _{3} $N can be a possible candidate for DMS at room temperature. Therefore, further experimental work would be required to confirm our theoretical prediction and propose a new DMS system.

\section{Acknowledgments}
GR acknowledges Higher Education Commission of Pakistan for supporting this research under the project electronic structure calculations using density functional theory. We also acknowledge GIK Institute for providing supercomputing facility. 

{}


\begin{thebibliography}{}


\bibitem{spintro2000} H. Ohno, \textit{et al.} Nature \textbf{408}, 944(2000).

\bibitem{spintro1} T. Dietl, Nat. Mater. {\bf 9}, 965 (2010).


\bibitem{spintro2} H. Ohno, Nat. Mater. {\bf 9}, 952 (2010).



\bibitem{TM1} F. Matsukura, H. Ohno,  and  T. Dietl, \textit{Handbook  of  Magnetic  Materials} {\bf 14},  edited  by  K.  H.  J.  Buschow,  Amsterdam:Elsevier, (2002).


\bibitem{TM2}T. Dietl,   Semicond. Sci. Technol. {\bf 17}, 377 (2002). 


\bibitem{TM3}T. Dietl, and H. Ohno,  MRS Bulletin {\bf 28}, 714 (2003). 



\bibitem{MCl1} J. H. Park, M. G. Kim, H. M. Jang, S. Ryu, Y. M. Kim, Appl. Phys. Lett. {\bf 84}, 1338 (2004).


\bibitem{MCl2} S. Zhou, K. Potzger, J. von Borany, R. Grotzschel, W. Skorupa, M. Helm, J. Fassbender, Phys. Rev. B {\bf 77},  035209 (2008).


\bibitem{g1} G. Rahman, V. M. Garc\'{i}a-Su\'{a}rez, and S. C. Hong, Phys. Rev. B {\bf 78}, 184404 (2008).


\bibitem{g2} G. Rahman, V. M. Garc\'{i}a-Su\'{a}rez, and J. M. Morbec, J. Magn. Magn. Mater. {\bf 328}, 104 (2013).


\bibitem{CaO} I. S. Elfimov, S. Yunoki, and G. A. Sawatzky, Phys. Rev. Lett. {\bf 89}, 216403 (2002).


\bibitem{SK1} S. K. Srivastava, P. Lejay, B. Barbara, S. Pailh\'{e}s, V. Madigou, and  G. Bouzerar, Phys. Rev. B {\bf 82}, 193203 (2010).


\bibitem{LE2} C. W. Zhang, and S. S. Yan, Appl. Phys. Lett. {\bf 95}, 232108 (2009). 


\bibitem{LE4} NU. Din and G. Rahman, RSC Adv. {\bf 4}, 29884 (2014).


\bibitem{LE5} G. Rahman, RSC Adv. {\bf 5}, 33674 (2015).

     
\bibitem{K.Li} K. Li, X. Du, Y. Yan, H. Wang, Q. Zhan, H. Jin, Phys. Lett. A {\bf 374}, 3671 (2010).
     

\bibitem{L.Shen} L. Shen, R. Q. Wu, H. Pan, G. W. Peng, M. Yang, Z. D. Sha, and Y. P. Feng, Phys. Rev. B  {\bf 78}, 073306 (2008).
     

\bibitem{H.Pan} H. Pan, J. B. Yi, L. Shen, R. Q. Wu, J. H. Yang, J. Y. Lin, Y. P. Feng, J. Ding, L. H. Van, and J. H. Yin, Phys. Rev. Lett. {\bf 99}, 127201 (2007).

\bibitem{G.R} G. Rahman, V. M. Garc\'{i}a-Su\'{a}rez, Appl. Phys. Lett. {\bf 96}, 052508 (2010).


\bibitem{S.W.Fan} S. W. Fan, K. L. Yao, and Z. L. Liu, Appl. Phys. Lett. {\bf 94}, 152506 (2009).

     
\bibitem{stb1} G. Rahman, NU. Din, V. M. Garc\'{i}a-Su\'{a}rez, E. Kan, Phys. Rev. B {\bf 87},  205205 (2013).


\bibitem{stb2} H. Pan, J. B. Yi, L.Shen, R. Q. Wu, J. H. Yang, J. Y. Lin, Y. P. Feng, J. Ding, L. H. Van, and J. H. Yin, Phys. Rev. Lett. {\bf 99}, 127201 (2007).




\bibitem{A.L}A. Lazicki, B. Maddox, W. J. Evans, C.-S Yoo, A. K. McMahan, W. E. Pickett, R. T. Scalettar, M. Y. Hu, and P. Chow, Phys. Rev. Lett. {\bf 95}, 165503 (2005).


\bibitem{A.L2} A. Lazicki, C. W. Yoo, W. J. Evans, M. Y. Hu, P. Chow, and W. E. Pickett, Phys. Rev. B {\bf 78}, 155133 (2008).



\bibitem{H} H. Schulz and K. H. Thiemann, Acta Crystallogr., Sect. A: Cryst. Phys., Diffr., Theor. Gen. Crystallogr. {\bf 35}, 309 (1979).


\bibitem{Hcap1} Y. H. Hu and E. Ruckenstein, J. Phys. Chem. A {\bf 107}, 9737 (2003).


\bibitem{Hcap2} Y. H. Hu and E. Ruckenstein, Ind. Eng. Chem. Res. {\bf 42}, 5135 (2003).




\bibitem{GaN} Y. Xie, Y. T. Qian, W. Z. Wang, S. Y. Zhang, and Y. H. Zhang, Science {\bf 272}, 1926 (1996).


\bibitem{LN5} P. Novak and F. R. Wagner, Phys. Rev. B {\bf 66}, 184434 (2002).


\bibitem{LN6} V. P. Antropov and V. N. Antonov, Phys. Rev. B {\bf 90}, 094406 (2014).


\bibitem{LN7} A. Jesche, L. Ke, J. L. Jacobs, B. Harmon, R. S. Houk, and P. C. Canfield, Phys. Rev. B {\bf 91}, 180403(R) (2015).

\bibitem{ostlin} A. \"{O}stlin, L. Chioncel, and E. Burzo, Romanian Journal of Physics \textbf{62}, 607 (2017). 

\bibitem{GR2017} G. Rahman, A. U. Rahman, S. Kanwal, and  P. Kratzer, EPL \textbf{119}, 57002 (2017).


\bibitem{H.J.Beister} H. J.  Beister, S. Haag \textit{et al} ; Angew. Chem, Int. Ed. Engl. {\bf 27}, 1101 (1988).


\bibitem{expRT} D. H. Gregory, P. M. \'{O}Meara, A. G. Gordon, J. P. Hodges, S. Short, and J. D. Jorgensen, Chem. Mater. {\bf 14}, 2063 (2002).





\bibitem{DFT} R. G. Parr and W. Yang, \textit{Density Functional Theory of Atoms and Molecules}, Oxford, New York, (1989).


\bibitem{HK} P. Hohenberg and W. Kohn, Phys. Rev. {\bf 136}, 3864 (1964).


\bibitem{siesta}  J. M. Soler, E. Artacho, J. D. Gale, A. Garc\'{i}a, J. Junquera, P. Ordej\'{o}n, and D. S\'{a}nchez-Portal, J. Phys. Condens. Matter {\bf 14}, 2745 (2002). 


\bibitem{lda} J. P. Perdew  and A. Zunger, Phys. Rev. B \textbf{23}, 5048 (1981).




\bibitem{Martin} N. Troullier and  Martıns, J. L. Phys Rev B  \textbf{43}, 1991 (1993).


\bibitem{gga} J. P. Perdew, K. Burke and M. Ernzerhof, Phys. Rev. Lett. {\bf 77}, 3865 (1996)


\bibitem{A.C} A. C. Ho, M. K. Granger, A. L. Ruoff, P. E. Van Camp, and V. E. Van Doren, Phys. Rev. B {\bf 59}, 6083 (1999).





\bibitem{Hug} A. Huq, J. W. Richardson, E.R. Maxey, D. Chandra, and W.-M. Chien, J.
Alloys Compd. {\bf 436}, 256 (2007)


\bibitem{Shun2009}S. Wu, Z. Dong, F. Boey, and P. Wu, Appl. Phys. Lett. \textbf{94}, 172104 (2009).


\bibitem{Sangster}J. Sangster and A. D. Pelton, J. Phase Equilib. 13, 291 (1992)


\bibitem{QE}Giannozzi, S. Baroni, N. Bonini, M. Calandara, R. Car, C. Cavazzoni, D.
Ceresoli, D. L. Chiarotti, M. Cococcioni, I. Dabo, A. Corso, S. de Dironcoli, S.
Fabris, G. Fratesi, R. Gebauer, U. Gerstmann, C. Gougoussis, A. Kokalj, M.
Lazzeri, L. Martin-Samos, N. Marzari, F. Mauri, R. Mazzarello, S. Paolini, A. Pasquarello, L. Paulatto, C. Sbraccia, S. Scandolo, G. Sclauzero, A.P. Setsonen, A. Smogunov, P. Umari, R.M. Wentzcovtch, J. Phys. Condense. Matter \textbf{21}, 39 (2009).


\bibitem {G.K} G. Kerker, Phys. Rev. B {\bf 23}, 6312 (1981).

\bibitem{expRT2} G. Nazri, Mater. Res. Soc. Symp. Proc.\textbf{135}, 117 (1989).


\bibitem{wu2011}A. Wu, Z. Dong, P. Wu, and F. Boey, J. Mater. Chem. {\bf 21}, 165
(2011).


\bibitem{wu2009}S. Wu, Z. Dong, F. Boey, and P. Wu, Appl. Phys. Lett. {\bf 94},
172104 (2009).




\bibitem{deltae}
When C$_{0}$-C$_{1}=3.63$\AA,
$E_{\rm FM}= -2J, \, E_{\rm AFM}=2J, \, \Rightarrow\,\Delta\,E=4J$. 
$E_{\rm FM}= -4J, \, E_{\rm AFM}=4J, \, \Rightarrow\,\Delta\,E=8J$, when C$_{0}$-C$_{2}=4.98$\AA. 

\end{thebibliography}
\end{document}